\documentstyle[prd,floats,aps]{revtex}

\begin{document}

\input{epsf.sty}

\draft

\twocolumn[\hsize\textwidth\columnwidth\hsize\csname
@twocolumnfalse\endcsname


\title{Galactic Collapse of Scalar Field Dark Matter}

\author{Miguel~Alcubierre$^{1}$,
F. Siddhartha~Guzm\'{a}n$^{1}$,
Tonatiuh~Matos$^{2}$,
Dar\'{\i}o~N\'{u}\~{n}ez$^{4,5}$,
L. Arturo~Ure\~{n}a-L\'{o}pez$^{2}$
and Petra~Wiederhold$^{3}$}

\address{
  $^{1}$Max-Planck-Institut f\"{u}r Gravitationsphysik, 
  Am M\"{u}hlenberg 1, D-14476 Golm, Germany. \\
  $^{2}$Departamento de F\'{\i}sica, $^{3}$Departamento de Control
  Autom\'atico, Centro de Investigaci\'{o}n y de Estudios Avanzados del
  IPN, A.P. 14-740, 07000 M\'{e}xico D.F., M\'{e}xico.\\
  $^{4}$ Center for Gravitational Physics and Geometry, Penn State
  University, University Park, PA 16802.\\
  $^{5}$Instituto de Ciencias Nucleares, 
  Universidad Nacional Aut{\'{o}}noma de M{\'{e}}xico, 
  A.P. 70-543, 04510 M{\'{e}}xico, D. F., Mexico.}

\date{\today; AEI-2001-123}

\maketitle


\begin{abstract}
  We present a scenario for galaxy formation based on the hypothesis
  of scalar field dark matter. We interpret galaxy formation through
  the collapse of a scalar field fluctuation.  We find that a $\cosh$
  potential for the self-interaction of the scalar field provides a
  reasonable scenario for galactic formation, which is in agreement
  with cosmological observations and phenomenological studies in
  galaxies.
\end{abstract}

\pacs{04.25.Dm, 
      95.30.Sf, 
      95.35.+d, 
      98.62.Ai, 
      98.80.-k} 

\vskip2pc]


\narrowtext   

In the last years, the quest concerning the nature of the dark matter
in the Universe has received much attention and has become of great
importance for understanding the structure formation in the Universe.
Some candidates for dark matter have been discarded and some others
have recently appeared. The standard candidates of the Cold Dark
Matter (CDM) model are axions and WIMP'S (Weakly Interacting Massive
Particles), which are themselves not free of problems. Axions are
massive scalar particles with no self interaction. In order for axions
to be an essential component of the dark matter content of the
Universe, their mass should be $m\sim 10^{-5}eV$. With this axion
mass, the scalar field collapses forming compact objects with masses
of order of $M_{crit}\sim 0.6m_{Pl}^{2}/m \sim
10^{-6}M_{\odot}$~\cite{seidel91,seidel94}, which corresponds to
objects with the mass of a planet.  Since the dark matter mass in
galaxies is ten times higher than the luminous matter, we would need
tenths of millions of such objects around the solar system, which is
clearly not the case. On the other hand, there are many viable
particles with nice features in super-symmetric theories, such as
WIMP'S.  However, since these candidates behave just like standard
CDM, they can not explain the observed scarcity of dwarf galaxies and
the smoothness of the galactic-core matter densities, since high
resolution numerical simulations with standard CDM predict an excess
of dwarf galaxies and density profiles with cusps~\cite{firmani}.
This is the reason why we need to look for alternative candidates that
can explain both the structure formation at cosmological level, the
observed amount of dwarf galaxies, and the dark matter density profile
in the core of galaxies.

In a recent series of papers, we have proposed that the dark matter in
the Universe is of a scalar field nature with a strong
self-interaction~\cite{DMCQG,SPHPRD,QSDMCQG,COSPRD,franky}. The scalar
field has been proposed as a viable candidate, since it mimics
standard CDM above galactic scales very well, reproducing most of the
features of the standard Lambda Cold Dark Matter ($\Lambda$CDM)
model~\cite{QSDMCQG,COSPRD,jeremy,peebles}.  However, at galactic
scales, the scalar field model presents some advantages over the
standard $\Lambda$CDM model.  For example, it can explain the observed
scarcity of dwarf galaxies since it produces a sharp cut-off in the
Mass Power Spectrum.  Also, its self-interaction can, in principle,
explain the smoothness of the energy density profile in the core of
galaxies~\cite{COSPRD,cross}. Nevertheless, the main problem when a
new dark matter candidate is proposed is the study of the final object
that would be formed as a result of a gravitational collapse.

The formation of galaxies through gravitational collapse of dark
matter is not an easy problem to understand. A good model for galaxy
formation has to take into account all the observed features of real
galaxies. For example, it seems that many disc galaxies contain a
black hole in their center, but others do not~\cite{Galx-bh}. Typical
galaxies are spiral, elliptical or dwarf galaxies (irregular galaxies
may be galaxies still evolving). In most spiral and elliptical
galaxies the luminous matter extends to $\sim 10-30$kpc, and the total
content of matter (including dark matter) is of the order of
$10^{10}-10^{12} M_{\odot}$, with about 10 times more dark matter than
luminous one.  The central density profile of the dark matter in
galaxies should not be cusp~\cite{Smooth}.  Even though the luminous
matter represents only a small fraction of the total amount of matter
in galaxies, it plays an important role in galaxy formation and
stability~\cite{DMCQG}. On the other hand, it is still not well
established if the mass of the central black hole and the mass of the
halo are correlated~\cite{correlated}, etc.

There are some ideas in this respect when dealing with a scalar field.
It is known that the final stage of a collapsed scalar field could be
a massive object made of scalar field particles in quantum coherent
states, like boson stars (for a complex scalar field) or oscillatons
(for a real scalar field)~\cite{seidel91,seidel94,luis}. It is thus
important to investigate whether the scalar field would collapse to
form structures of the size of galaxies and provide the correct
properties of any galactic dark matter candidate, like growing
rotation curves and appropriate dark matter distribution functions.
Also, we need to know which are the conditions that must be imposed on
the scalar field particles.

Our main aim in this paper is to present a plausible scenario for
galaxy formation under the scalar field dark matter (SFDM) hypothesis.
Through a gravitational cooling process~\cite{seidel91,seidel94}, a
cosmological fluctuation of the scalar field collapses to form a
compact oscillaton by ejecting part of the field. The key idea
consists precisely in assuming that such final object could distribute
as galactic dark matter does. The final configuration then should
consist of a central object (a core), i.e. an oscillaton, surrounded
by a diffuse cloud of scalar field, both formed at the same time due
to the same collapse process.

At the cosmological scale, it is found that the mass of the boson is
not the only parameter that determines the power spectrum.  The
self-interaction of the scalar field is also important. Following an
analogous procedure to the one used in particle physics, we may write
a phenomenological Lagrangian with all the terms we need in order to
reproduce the observed Universe. In particular, if one uses a
minimally coupled real scalar field $\Phi$ with a self-interaction
potential of the form ($\kappa _{0}=8\pi G$, we use natural units
such that $\hbar=c=1$)
\begin{equation}
V(\Phi ) = V_{0} \left[ \cosh {(\lambda \sqrt{\kappa_{0}}\Phi )-1}
\right] \, ,
\label{cosh}
\end{equation}
then one can show~\cite{QSDMCQG,COSPRD} that the SFDM model reproduces
well all the successes of the standard $\Lambda$CDM above galactic
scales. The free parameters of the scalar potential, $V_{0}$ and the
scalar field mass $m_\Phi=\lambda \sqrt{V_0 \kappa_{0}}$, can be
fitted by cosmological observations.  Doing this one finds
that~\cite{COSPRD} (see also~\cite{Hu})
\begin{eqnarray}
\lambda &\simeq & 20.28\, , 
\label{lambda} \\
V_0 &\simeq &(3\times 10^{-27}\,m_{Pl})^{4}  \, ,
\label{V0} \\
m_\Phi &\simeq &9.1\times 10^{-52}\,m_{Pl}=1.1\times 10^{-23}eV \, ,
\label{mphi}
\end{eqnarray}
with $m_{Pl} \equiv G^{-1/2}$ the Planck mass.

If a galaxy is an oscillaton, i.e. an oscillating soliton object, it
must correspond to coherent scalar oscillations around the minimum of the
scalar potential~(\ref{cosh}). The scalar field $\Phi$ and the metric
coefficients (considering the spherically symmetric case) would be
time dependent, and it has been shown that such a configuration can be
stable, non-singular and asymptotically flat~\cite{seidel91}. For the
scalar field collapse, the critical value for the mass of an
oscillaton (the maximum mass for which a stable configuration
exists) will depend on the mass of the boson. Roughly speaking, if we
take $m_{\Phi }=1.1\times 10^{-23}eV$, and use the formula for the
critical mass of the oscillaton corresponding to a scalar field with a
$\Phi^2$ potential (i.e. just a mass term), we expect the critical
mass to be~\cite{seidel91,seidel94}
\begin{equation}
M_{crit}\sim 0.6 \, \frac{m_{Pl}^{2}}{m_\Phi} \sim 10^{12}M_{\odot} \, .
\label{masa}
\end{equation}
This is a surprising result: the critical mass of the model shown
in~\cite{QSDMCQG,COSPRD} is of the order of magnitude of the dark
matter content of a standard galactic halo.

In order to study this situation for the case of a potential of the
form~(\ref{cosh}), we present a numerical simulation of Einstein's
equations in which the energy momentum tensor is that of a real scalar
field. The scenario of galactic formation we assume is as follows: a
sea of scalar field particles fills the Universe and forms localized
primordial fluctuations that could collapse to form stable objects,
which we will interpret as the dark matter halos of galaxies.

We evolve the spherically symmetric line element
\begin{equation}
ds^2 = -\alpha^2(r,t) dt^2 + a^2(r,t) dr^2 + r^2 d\Omega^2 \, ,
\label{metrica}
\end{equation}
with $\alpha(r,t)$ the lapse function and $a(r,t)$ the radial metric
function.  We choose the polar-areal slicing condition (i.e. we force
the line element to have the above form at all times, so that the area
of a sphere with $r=R$ is always equal to $4 \pi R^2$).  This choice
of gauge will force the lapse function $\alpha(r,t)$ to satisfy an
ordinary differential equation in $r$ (see below).

The energy momentum tensor of the scalar field is
\begin{equation}
T_{\mu \nu }=\Phi_{,\mu }\Phi_{,\nu} - \frac{g_{\mu \nu}}{2} \,
\left[ \Phi^{,\alpha }\Phi_{,\alpha} + 2V(\Phi) \right] \, .
\end{equation}
We now introduce the first order variables $\Psi =\Phi_{,r}$ and
\mbox{$\Pi=a \Phi_{,t} / \alpha$}.  Using these new variables, the
Hamiltonian constraint becomes 
\begin{equation}
\frac{a_{,r}}{a} = \frac{1-a^{2}}{2r}+\frac{\kappa_{0}r}{4}
\left[ \Psi^{2} + \Pi^{2} + 2a^{2}V\right] \, ,
\label{ham}
\end{equation}
and the polar-areal slicing condition takes the form:
\begin{eqnarray}
\frac{\alpha _{,r}}{\alpha } &=&\frac{a_{,r}}{a} + \frac{a^{2}-1}{r}
- \kappa_{0} r a^{2} V  \, .
\label{slice}
\end{eqnarray}
All other components of Einstein's equations either vanish, or are a
consequence of the last two equations.
 
The Klein-Gordon (KG) equation now reads
\begin{eqnarray}
\Phi_{,t} &=& \frac{\alpha}{a} \, \Pi \, ,
\label{KG1} \\
\Pi_{,t} &=& \frac{1}{r^{2}}\left( \frac{r^2 \alpha \Psi}{a}\right)
_{,r} - a\alpha \, \frac{dV}{d\Phi } \, ,
\label{KG2} \\
\Psi_{,t} &=& \left( \frac{\alpha \Pi}{a} \right)_{,r} \, .
\label{KG3}
\end{eqnarray}

Equations (\ref{ham}-\ref{KG3}) form the complete set of
differential equations to be solved numerically.  For numerical
purposes, the evolution equation for $\Pi$ above is further
transformed into the equivalent form:
\begin{equation}
\Pi_{,t} = 3 \frac{d}{dr^3} \left( \frac{r^2 \alpha \Psi}{a} \right)
- a\alpha \, \frac{dV}{d\Phi } \, .
\end{equation}
Notice that the first term on the right hand side of this equation
includes now a first derivative with respect to $r^3$ (and not a third
derivative).  The reason for doing this transformation has to do with
the numerical regularization near the origin of the $1/r^2$ factor in
equation~(\ref{KG2}) above (see Ref.~\cite{Hawley}).

In order to deal with non-dimensional units, we define $x=l \, r$. A
natural scale for the potential is given by $l^{-1}=1/\sqrt{\kappa_{0}
  V_{0}}=12 \, pc = 40 \, yr$. The parameter $l$ also gives the time
scale $\tau = t l$. The scalar field has an initial Gaussian profile
\begin{equation}
\sqrt{\kappa_{0}}\, \Phi (x,t=0) = A e^{-x^{2}/s^{2}} \, ,
\label{gauss}
\end{equation}
with $A$ the amplitude and $s$ the width of the Gaussian. The physical
properties describing the state of the system are the energy density
of the scalar field $\rho_{\Phi }=(m_{Pl}^{2}l^{2}/8\pi) \, \rho_{s}$,
with $\rho _{s}$ the dimensionless quantity
\begin{equation}
\rho_s =\frac{1}{2a^{2}} \left( \Psi ^{2}+\Pi ^{2} \right) + V \, ,
\label{densidad}
\end{equation}
and the integrated mass
\begin{equation}
M(x) = \frac{m_{Pl}^{2}}{2 l}\int_{0}^{x}\rho_{s}(X)\,X^{2}dX \, .
\label{mass}
\end{equation}

Some details about our numerical implementation are in order. To
integrate the KG equation numerically we use a method of lines with
standard centered second order finite differences in space, and a
third order in time integrator.  At the outer boundary we impose a
condition for simple outgoing radial waves.  We deal with the
singularity at $x=0$ by straddling the origin and imposing the
adequate parity conditions for each function on an auxiliary point at
$x=-\Delta x/2$. The ordinary differential equations for $a$ and
$\alpha$ are solved using a second order Runge-Kutta method.  The third
order in time integration has been chosen to reduce as much as
possible the numerical dissipation in our code, which we have found to
be crucial in order to obtain reliable results for the long time runs
we have studied.

The numerical simulations suggest that the critical mass for the case
considered here, using the scalar potential~(\ref{cosh}), is
approximately~\cite{luis,futuro}
\begin{equation}
M_{crit}\simeq 0.1 \frac{m_{Pl}^{2}}{\sqrt{\kappa_0 V_0}} =
2.5 \times 10^{13}M_{\odot} \, .
\label{masa1}
\end{equation}

\begin{figure}[ht]
\centerline{ \epsfysize=6cm \epsfbox{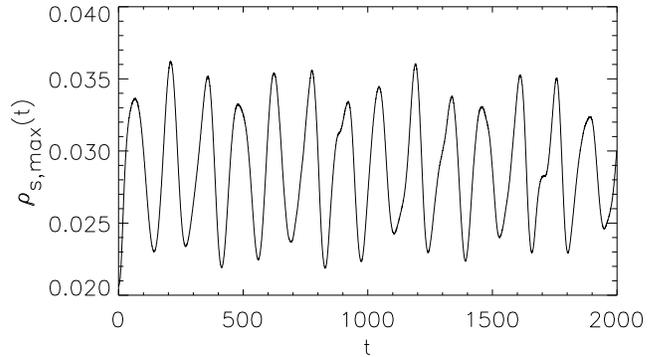}}
\caption{Temporal evolution of the maximum value of the energy
  density. The parameters for the initial configuration are $A=0.01,\,
  s=2.0$ (see~(\ref{gauss})), which correspond to an initial mass of
  $M_i = 3.25 \times 10^{12} \, M_{\odot}$. See also
  Fig.~\ref{fig:params3}.}
\label{fig:params}
\end{figure}

The results of the numerical simulations are as follows. Essentially,
we have found three different types of behavior for the scalar field
collapse.  In the first case, a generic feature is that scalar field
distributions with an initial mass slightly larger than the critical
mass collapse very violently and form a black hole.  In the second
type of behavior, fluctuations with an initial mass significantly
smaller than the critical mass can not form stable oscillatons: the
scalar field is completely ejected out as the system
evolves~\cite{futuro}.  The third behavior corresponds to a
case where a fraction of the initial density is spread out, leaving
an oscillating object that appears to be stable. This situation
happens in a narrow window of initial conditions, between $0.05 - 1
\times M_{crit}$ ~\cite{futuro}.

In Fig.~\ref{fig:params} we show the evolution in time of the maximum
value of the energy density for an initial configuration that results
in the formation of an oscillaton.  In this case, we have taken an
initial configuration with $A=0.01$, $s=2.0$, which implies an initial
mass of \mbox{$M_i = 3.25 \times 10^{12} \, M_{\odot}$}.  For this run
we used $\Delta x = 0.005$, $\Delta t = \Delta x/10$, and
10,000 grid points, which puts the outer boundaries at $x=50$.  The
run was followed until $t=2000$, some 40 light crossing times.

\begin{figure}[ht]
\centerline{ \epsfysize=6cm \epsfbox{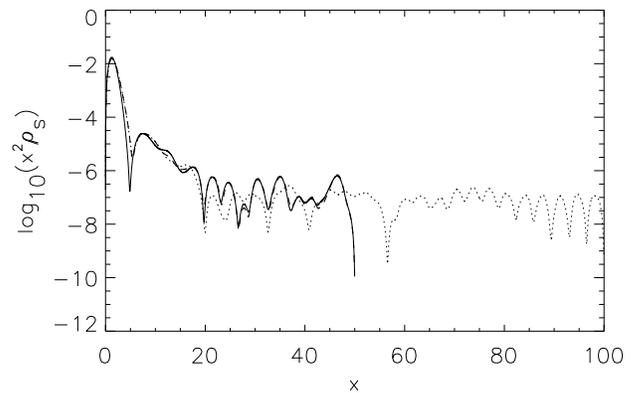}}
\caption{Energy density of the object at $t=2000$.
  In the plot we show $x^2 \rho_s (x)$ using a logarithmic scale.  The
  solid line corresponds to a run with $\Delta x=0.005$ and the
  boundaries at $x=50$, the dashed line to a run with $\Delta x =
  0.01$ and boundaries at $x=50$, and the dotted line to a run with
  $\Delta x = 0.01$ and boundaries at $x=100$.}
\label{fig:params2}
\end{figure}

In Fig.~\ref{fig:params2} we plot the energy density $x^2 \rho_s (x)$
of the object at $t=2000$, using a logarithmic scale, for three
different runs: the run mentioned above with $\Delta x = 0.005$ and
boundaries at $x=50$, and two runs with half the resolution ($\Delta x
= 0.01$) and with the boundaries at $x=50$ and $x=100$ respectively.
Notice how the the two runs with the boundaries at the same location
coincide very well throughout the computational domain.  The run with
the boundaries twice as far agrees well with the other runs for
$x<20$, but differs significantly outside where $x^2\rho_s < 10^{-6}$.
This indicates that at such low levels, the solution is dominated by
boundary effects.  Our boundary condition is clearly introducing
numerical noise at such late times.

\begin{figure}[ht]
\centerline{ \epsfysize=6cm \epsfbox{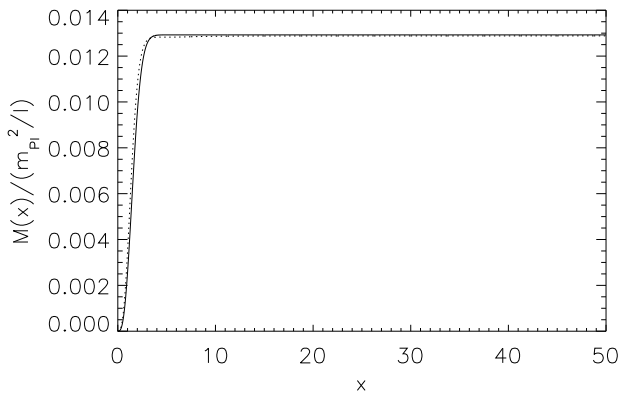}}
\caption{Initial (solid line) and final (dotted line) integrated
  masses.}
\label{fig:params3}
\end{figure}

Figure~\ref{fig:params3} shows the integrated mass~(\ref{mass}) for
the initial and final ($t=2000$) stages of the evolution.  A small
drop of $\sim 0.5\%$ in the total integrated mass can be observed, but
convergence tests suggest that most (if not all) of this mass loss is
caused by a small amount of numerical dissipation still present in our
numerical method.  This implies that the system does not radiate any
significant amount of energy during the time of the simulation, which
indicates that the object is very stable.

From the cosmological point of view, the narrow window of initial
conditions means that not all fluctuations will collapse into stable
objects. Moreover, the collapsed objects will have masses of the same
order of magnitude $M_{\rm final}\sim 10^{12}M_{\odot}$, as it seems
to be precisely the case for galaxies.
 
Summarizing, from the results of the numerical simulations of the
collapse of the real scalar field with a $\cosh$ potential we find
many similarities with the structure of the halos of galaxies. The
scalar field density profile is not singular at the center. This fact,
and the values of the final masses obtained using the cosmological
values~(\ref{V0}) and~(\ref{mphi}) for the parameters of the
self-interaction potential, could correspond to objects like realistic
galaxies.  Moreover, it is in agreement with the observational
constraints related to the phenomenological maximum galactic mass
pointed out by Salucci and Burkert~\cite{salbur}.  Therefore, we
expect that fluctuations of this scalar field, due to Jeans
instabilities, will in general collapse to form objects of the order
of the mass of the halo of a typical galaxy.

We have shown before~\cite{QSDMCQG,COSPRD} that the SFDM model could
be a good model for the universe at cosmological level, here we see
that the scalar field could also be a good candidate for the dark
matter content of individual galaxies (as suggested
in~\cite{DMCQG,SPHPRD,franky}).


\acknowledgements{We would like to thank Erasmo G\'omez, Aurelio
  Esp\'{\i}ritu and Kenneth Smith for technical support, and Edward
  Seidel for useful discussions. This work was partly supported by
  CONACyT M\'exico, under grants 119259 (L.A.U.), 32138-E and 34407-E. DN
  acknowledges the DGAPA-UNAM grant IN-121298.}


\end{document}